\pdfoutput=1
\documentclass[12pt]{article}

\usepackage{amsmath,graphicx}
\usepackage{simplewick}
\usepackage{epsf}
\usepackage{graphicx,epsfig}
\usepackage{amsfonts}
\usepackage{amssymb}
\usepackage[nosort]{cite}
\usepackage{setspace}

\newcommand{\nc}{\newcommand}
\nc{\ba}{\begin{eqnarray}}
\nc{\ea}{\end{eqnarray}}

\def\bk{{\bf k}}
\def\bp{{\bf p}}
\def\bq{{\bf q}}
\def\bx{{\bf x}}
\def\by{{\bf y}}

\def\CH{{\cal H}}
\def\CL{{\cal L}}

\def\Mpl{M_{\rm P}}

\makeatletter
\renewcommand\section{\@startsection {section}{1}{\z@}%
                                 {-3.5ex \@plus -1ex \@minus -.2ex}%
                                   {2.3ex \@plus.2ex}%
                                   {\normalfont\large\bfseries}}
\renewcommand\subsection{\@startsection{subsection}{2}{\z@}%
                                   {-3.25ex\@plus -1ex \@minus -.2ex}%
                                     {1.5ex \@plus .2ex}%
                                     {\normalfont\bfseries}}
\renewcommand\subsubsection{\@startsection{subsubsection}{3}{\z@}%
                                   {-3.25ex\@plus -1ex \@minus -.2ex}%
                                     {1.5ex \@plus .2ex}%
                                     {\normalfont\itshape}}
\makeatother

\newcommand{\Letter}{
\setlength{\textwidth}{16.5cm}
   \setlength{\textheight}{23cm}
    \hoffset=-0.5in
\voffset=-2.1cm }

\Letter

\begin{document}
\newcommand{\be}{\begin{equation}}
\newcommand{\ee}{\end{equation}}
\newcommand{\bea}{\begin{eqnarray}}
\newcommand{\eea}{\end{eqnarray}}
\newcommand{\barr}{\begin{array}}
\newcommand{\earr}{\end{array}}
\def\bal#1\eal{\begin{align}#1\end{align}}

\thispagestyle{empty}
\begin{flushright}
%\parbox[t]{1.5in}{hep-th/yymmnnn}
\end{flushright}

\vspace*{0.3in}
\begin{spacing}{1.1}

\begin{center}
{\large \bf On the equation-of-motion versus in-in approach \\
in cosmological perturbation theory}

\vspace*{0.3in} {Xingang Chen$^1$, Mohammad Hossein Namjoo$^1$, Yi Wang$^2$}
\\[.3in]
{\em
$^1$Department of Physics, The University of Texas at Dallas, Richardson, TX 75083, USA \\
$^2$Department of Physics, The Hong Kong University of Science and Technology,\\
Clear Water Bay, Kowloon, Hong Kong, P.R.China} \\[0.3in]

\end{center}

\begin{center}
{\bf
Abstract}
\end{center}
\noindent
In this paper, we study several issues in the linear equation-of-motion (EoM) and in-in approaches of computing the two-point correlation functions in multi-field inflation. We prove the equivalence between this EoM approach and the first-principle in-in formalism. We check this equivalence using several explicit examples, including cases with scale-invariant corrections and scale-dependent features. Motivated by the explicit proof, we show that the usual procedures in these approaches can be extended and applied to some interesting model categories beyond what has been studied in the literature so far. These include the density perturbations with strong couplings and correlated multi-field initial states.

\vfill

\newpage
\setcounter{page}{1}

%\tableofcontents

\newpage

\section{Introduction}

The cosmic inflation \cite{Guth:1980zm,Linde:1981mu,Albrecht:1982wi,Starobinsky:1980te,Sato:1980yn} is a unique probe of high energy physics, providing access to energy scales that is inaccessible in colliders in the foreseeable future. The most important and well measured observable from inflation is the power spectrum of the primordial density fluctuations. The current observations can be well fit by a nearly scale invariant power spectrum with a slightly red tilt \cite{Planck:2015xua,Ade:2015lrj}. The simple form of the primordial power spectrum will be tested or challenged by many on-going and future experiments. For example, future ground based telescopes can constrain the CMB polarization with a precision close to cosmic variance; future 21~cm experiments may push the measurement of primordial power spectrum to unprecedented precision \cite{Furlanetto:2006jb}, where the limitation of cosmic variance is as small as a relative error of $10^{-10}$.

The exciting future of cosmological experiments poses challenges for the computational methods of inflationary perturbations. The best established method to calculate the inflationary perturbations is the in-in formalism \cite{Weinberg:2005vy,Chen:2010xka,Wang:2013zva}. It is an operator approach derived from first-principles, and is defined to compute the correlation functions to all orders in perturbations. However, sometimes in practice, it is increasingly hard to compute high order corrections in the in-in formalism, because of the algebraic complexity, the spontaneously breaking of Lorentz invariance, and the subtleties of UV and IR divergences \cite{Chen:2009we, Chen:2009zp}. Moreover, non-perturbatively, the in-in formalism is not well defined, and in some situations provides formal expressions at most.

Alternatively, solving the equations of motion (EoM) for fields is another way to compute the primordial perturbations.
In contrast to the in-in formalism, the method of computing correlation functions to all orders in the EoM approach is not known, although there is a proposal on how to generate all the tree-level results \cite{Weinberg:2008mc}. The method of linearly computing the two-point correlation functions in the multi-field inflation models, however, is well known in the literature. See e.g.~\cite{GrootNibbelink:2001qt} for earlier works, \cite{Weinberg:2008zzc} for a review, and \cite{Gao:2013ota,Dias:2015rca} for recent examples.
Despite of the limitation, in the EoM approach, we deal with differential equations of c-numbers.
This is particularly useful for numerical computations. In addition, as we will show, this method is even defined non-perturbatively at the linear level.

The EoM approach used to linearly compute the two-point correlation functions has several special but well-known steps.
Despite the general belief, an explicit and general proof for the equivalence between this EoM approach and the in-in formalism is not present. The equivalence has only been demonstrated in some special examples, e.g~\cite{Gao:2013ota,AkbarUnpub}.
In this paper we provide such a proof. Inspired by the explicit proof, we also discuss several extensions beyond the known formalisms.

This paper is organized as follows.
In Sec.~\ref{Sec:Review}, we first review the procedures used to compute the power spectrum in multi-field inflation models in both the in-in formalism and the EoM approach.
In Sec.~\ref{Sec:Proof}, We prove the equivalence between the EoM approach and the in-in formalism at the linear order.
In Sec.~\ref{Sec:Examples}, we use explicit multi-field inflationary models to check this equivalence. Motivated by the explicit proof of the equivalence in Sec.~\ref{Sec:Proof}, in Sec.~\ref{Sec:EoM_special} we show that the usual procedures in these approaches can be extended and applied to some new categories of model examples. These include the cases where the couplings are non-perturbative and where the initial states are correlated multiple fields.
We conclude in Sec.~\ref{Sec:conclusion}.

\section{Tree-level power spectrum in multi-field inflation models}
\label{Sec:Review}
\setcounter{equation}{0}

In this section, we outline the two approaches of computing the two-point correlation functions in multi-field inflation models, namely linearly solving the EoM (Sec.~\ref{Sec:EoM}) and computing tree-level diagrams in the in-in formalism (Sec.~\ref{Sec:inin}).
We emphasize some seemingly ad hoc procedures required in the EoM approach.

\subsection{EoM approach}
\label{Sec:EoM}

Consider a multifield inflation model with $N$ number of fields, $\phi_a(\bx,t)$, $a=1,\dots,N$.
Denote $\bar\phi_a(\bx,t)$ and $\dot{\bar\phi}_a(\bx,t)$ as the background solution, and $\delta\phi_a(\bx,t)$ and $\delta\dot\phi_a(\bx,t)$ as the perturbations.
Denote $\tilde L$ as the part of Lagrangian that is quadratic or higher in perturbations $\delta\phi_a$ and $\delta\dot\phi_a$,
\begin{align}
\tilde L (\delta\phi_a, \delta\dot\phi_a, t)
\equiv L (\phi_a,\dot\phi_a)
%\nonumber \\
- L (\bar\phi_a, \dot{\bar\phi}_a)
- \int d^3x \frac{\partial L}{\partial \bar\phi_a} \delta\phi_a
- \int d^3x \frac{\partial L}{\partial \dot{\bar\phi}_a} \delta\dot\phi_a ~.
\label{tL_def}
\end{align}
The EoM is
\bal
\frac{d}{dt} \left( \frac{\partial \tilde L}{\partial \delta\dot\phi_a} \right) - \frac{\partial\tilde L}{\partial\delta\phi_a} =0 ~,
\quad
a=1,\dots,N ~.
\label{EoM}
\eal
Here the partial derivatives are the functional derivatives. To linearly compute the two-point correlation functions, we only keep the quadratic terms in ${\tilde L}$. So (\ref{EoM}) is $N$ number of coupled second order linear differential equations.
The mode function of $\delta\phi_a$, denoted as $u_a$, is defined in the momentum space,
\bea
u_a(\bk,t) = \int d^3x \delta\phi_a(\bx,t) e^{i\bk\cdot\bx} ~.
\eea
These mode functions satisfy the same EoM (\ref{EoM}) in the momentum space.

In this section, we consider the usual cases where the fields are decoupled initially. Consequently, each field takes the following initial conditions at the initial time $t_0$,
\bea
u_a (\bk, t_0) = u^{\rm ini}_a (\bk) ~,
\quad
\dot u_a (\bk, t_0) = \tilde u^{\rm ini}_a (\bk) ~,
\quad
a=1,\dots,N ~,
\label{Initial_Cond}
\eea
where the $u_a^{\rm ini}$ and $\tilde u_a^{\rm ini}$ satisfy the initial commutation condition. For example, if the fields have canonical kinetic terms, these conditions are
\bea
a^3 (t_0) \left( u_a \tilde u_a^* - {\rm c.c.} \right) = i ~,
\quad ({\rm no~sum~over~{\it a}}) ~.
\label{Initial_W-cond}
\eea
We make a few clarification remarks on the terminologies ``the initial commutation relation" used here and the ``Wronskian condition". Wronskian condition is a time dependent condition for a decoupled single field mode, while the initial commutation relation is specified only at the initial time $t_0$. If the fields are decoupled initially, the initial commutation relation is the Wronskian condition evaluated at $t_0$. These conditions apply to both the Bunch-Davies vacuum and non-Bunch-Davies vacua, therefore in (\ref{Initial_Cond}) the values of $u_a^{\rm ini}$ and $\tilde u_a^{\rm ini}$ can be chosen arbitrarily as long as (\ref{Initial_W-cond}) holds.
After $t_0$, we turn on interaction, so (\ref{Initial_W-cond}) is only needed at $t_0$ but in general may not be satisfied for all $t$.

The procedure of getting the two-point correlation functions is as follows. We need to solve the EoM (\ref{EoM}) for {\em $N$ times separately} with different sets of initial conditions.

For example, for the 1st time, the initial condition is taken to be
\bea
u^{(1)}_1 = u^{\rm ini}_1 ~,
\quad
\dot u^{(1)}_1 = \tilde u^{\rm ini}_1 ~;
\quad
u^{(1)}_a =0 ~,
\quad
u^{(1)}_a =0
\quad (a\ne 1) ~.
\label{initial_EoM_1}
\eea
This leads to the 1st solution at the final time $t$, which we denote as $u^{(1)}_a(\bk,t)$, $a=1,\dots,N$.
For the 2nd time, the initial condition is taken to be
\bea
u^{(2)}_2 = u^{\rm ini}_2 ~,
\quad
\dot u^{(2)}_2 = \tilde u^{\rm ini}_2 ~;
\quad
u^{(2)}_a =0 ~,
\quad
\dot u^{(2)}_a =0
\quad
(a\ne 2) ~.
\label{initial_EoM_2}
\eea
This leads to the 2nd solution at the final time $t$, $u^{(2)}_a(\bk,t)$, $a=1,\dots,N$.
Repeat this for $N$ times, and finally we get $N$ sets of solutions at $t$,
$u^{(\alpha)}_a(\bk,t)$, where $\alpha, a=1,\dots,N$.

The two-point correlation functions of the fields are determined by these solutions,
\bea
\langle \delta\phi_{a\bp}(t) \delta\phi_{b\bq}(t) \rangle
= (2\pi)^3 \sum_{\alpha=1}^{N} u^{(\alpha)}_a(\bp,t) u^{(\alpha)*}_b(\bq,t) \delta^3(\bp+\bq) ~.
\label{2pt_EoM}
\eea

Notice that, if we directly supply all the initial conditions (\ref{Initial_Cond}) to the EoM (\ref{EoM}) and solve it only for once, we in general would have got the wrong results, even though the initial conditions are chosen such that the multiple fields are decoupled and the Lagrangian contains no interaction initially.

\subsection{In-in formalism}
\label{Sec:inin}

Similarly, we denote $\bar\phi_a(\bx,t)$ and $\bar\pi_a(\bx,t)$ as the background solutions, and $\delta\phi_a(\bx,t)$ and $\delta\pi_a(\bx,t)$ as the perturbations, where $\bar\pi_a(\bx,t)$ and $\delta\pi_a(\bx,t)$ are conjugate momenta associated with the fields  $\bar\phi_a(\bx,t)$ and $\delta\phi_a(\bx,t)$, respectively.  We also denote $\tilde H$ as the part of Hamiltonian that is quadratic or higher in perturbations $\delta\phi_a$ and $\delta\pi_a$,
\begin{align}
\tilde H \left[ \delta\phi_a,\delta\pi_a, t \right] \equiv
H\left[ \phi_a,\pi_a \right] - H \left[ \bar\phi_a,\bar\pi_a \right]
- \sum_a \int d^3x \frac{\partial \CH}{\partial \bar\phi_a} \delta\phi_a
- \sum_a \int d^3x \frac{\partial \CH}{\partial \bar\pi_a} \delta\pi_a ~.
\label{tildeH}
\end{align}
To get the tree-level two-point correlation functions, we only keep the quadratic terms in $\tilde H$.
The evolution of the fields is governed by the following EoM,
\begin{align}
\delta\dot\phi_a(\bx,t) = i \left[ \tilde H,\delta\phi_a(\bx,t) \right] ~,
\quad
\delta\dot\pi_a(\bx,t) = i \left[ \tilde H,\delta\pi_a(\bx,t) \right] ~,
\label{EoM_Hamiltonian}
\end{align}
$a=1,\dots,N$.
It is easy to show \cite{Chen:2010xka} that this EoM is equivalent to (\ref{EoM}).

The in-in formalism is a perturbative method of solving the EoM (\ref{EoM_Hamiltonian}).
We split $\tilde H$ into two parts,
\bea
\tilde H \left[ \delta\phi_a,\delta\pi_a,t\right]
= H_0 \left[ \delta\phi_a,\delta\pi_a,t\right]
+ H_I \left[ \delta\phi_a,\delta\pi_a,t\right] ~,
\label{Hsplit}
\eea
where $H_0$ is the quadratic kinematic part of the Hamiltonian describing the leading order $N$ number non-interacting fields. The fields with this leading order behavior, denoted as $\delta\phi_a^I$ and $\delta\pi_a^I$, are called the interaction picture fields and follow from
\begin{align}
\delta\dot\phi^I_a = i \left[ H_0,\delta\phi^I_a \right] ~,
\quad
\delta\dot\pi^I_a = i \left[ H_0,\delta\pi^I_a \right] ~.
\label{EoM_Interaction_Pic}
\end{align}
The initial conditions for this EoM is given by (\ref{Initial_Cond}) and (\ref{Initial_W-cond}). Because interaction is not introduced in $H_0$, the mode function of each interaction picture field satisfies the Wronskian condition.

The correlation function $\langle Q \rangle$ is then given by perturbatively expanding the following expression in terms of powers of $H_I$,
\begin{align}
\langle Q \rangle
= \langle \Omega| \left[ \bar T \exp \left( i \int_{t_0}^t H_I(t) dt \right) \right]
~Q^I(t)~
\left[ T \exp \left( -i \int_{t_0}^t H_I(t) dt \right)\right] |\Omega\rangle ~,
\label{in-in_formalism}
\end{align}
where
\begin{align}
H_I(t) &\equiv H_I\left[ \delta\phi^I_a, \delta\pi^I_a, t\right] ~,
\\
Q^I(t) &\equiv Q\left[\delta\phi^I_a,\delta\pi^I_a\right] ~.
\end{align}
In the perturbative expansion, we plug in solutions of (\ref{EoM_Interaction_Pic}) in terms of the following decomposition,
\begin{align}
\delta\phi_a &= u_a a_{a,\bk} + u_a^* a_{a,-\bk}^\dagger ~,
\nonumber \\
\delta\pi_a &= w_{a} a_{a,\bk} + w_{a}^* a_{a,-\bk}^\dagger ~,
\quad
{\rm (no~sum~over~{\it a})~, }
\label{phipi_decomp_usual}
\end{align}
where $a=1,\dots,N$ labels different fields. Note that each field is expanded into one set of the creation and annihilation operators.\footnote{This is the decomposition that has been used so far \cite{Weinberg:2005vy,Chen:2010xka,Wang:2013zva}. In Sec.~\ref{Sec:correlated_initial}, we discuss the extension.}

The in-in formalism is derived from the first-principles \cite{Weinberg:2005vy,Chen:2010xka,Wang:2013zva}, while the status of the EoM approach is not entirely clear given the procedures.
For single field inflation, the equivalence of the above two approaches is obvious since they are solving the equivalent EoM, (\ref{EoM}) and (\ref{EoM_Hamiltonian}). For multifield inflation, it is not obvious why we should solve the EoM separately for $N$ times, instead of solving it once with all the initial conditions, even though in the model setup we are considering the cases where the initial fields are decoupled and non-interacting. In the following section, we justify this procedure in the EoM approach by showing that it satisfies the same set of first-principles as in the in-in formalism.

\section{The proof of equivalence}
\label{Sec:Proof}
\setcounter{equation}{0}

For completeness, let us start by briefly repeating the setup using a more detailed notation following \cite{Weinberg:2005vy,Chen:2010xka,Wang:2013zva}.
Denote the Hamiltonian of the system as
\bea
H[\phi(t),\pi(t)] \equiv
\int d^3x \CH [ \phi_a(\bx,t),\pi_b(\bx,t)] ~,
\eea
where $a,b=1,...,N$ label the number of fields.
We perturb $\phi_a$ and $\pi_a$ around a time-dependent background, $\bar\phi_a(\bx,t)$ and $\bar\pi_a(\bx,t)$,
\begin{align}
\phi_a(\bx,t) = \bar\phi_a(\bx,t) + \delta\phi_a(\bx,t) ~,
\quad
\pi_a(\bx,t) = \bar\pi_a(\bx,t) + \delta\pi_a(\bx,t) ~.
\end{align}
The background satisfies the classical equations of motion,
\bea
\dot{\bar\phi}_a(\bx,t) = \frac{\partial \CH}{\partial \bar\pi_a} ~,
\quad
\dot{\bar\pi}_a(\bx,t) = - \frac{\partial \CH}{\partial \bar\phi_a} ~.
\label{bkgd_eom}
\eea
We expand the Hamiltonian as
\begin{align}
H\left[ \phi(t),\pi(t) \right] =&~ H \left[ \bar\phi(t),\bar\pi(t) \right]
+ \sum_a \int d^3x \frac{\partial \CH}{\partial \bar\phi_a(\bx,t)} \delta\phi_a(\bx,t)
+ \sum_a \int d^3x \frac{\partial \CH}{\partial \bar\pi_a(\bx,t)} \delta\pi_a(\bx,t)
\nonumber \\
& + \tilde H \left[ \delta\phi(t),\delta\pi(t);t \right] ~,
\label{Hexpand}
\end{align}
where we use $\tilde H$ to denote terms of quadratic and higher orders in perturbations.

If we impose the canonical commutation relation for $\phi_a$ and $\pi_a$, we immediately get the same relation for $\delta\phi_a$ and $\delta\pi_a$,
\begin{align}
& \left[ \delta\phi_a(\bx,t),\delta\pi_b(\by,t)\right] = i\delta_{ab} \delta^3(\bx-\by) ~,
\nonumber \\
& \left[ \delta\phi_a(\bx,t),\delta\phi_b(\by,t)\right] =
\left[ \delta\pi_a(\bx,t),\delta\pi_b(\by,t) \right] =0 ~.
\label{comm_relation_0}
\end{align}
Using these relations, we can derive the equations that govern the evolution of the perturbations
\begin{align}
\delta\dot\phi_a(\bx,t) = i \left[ \tilde H \left[ \delta\phi(t),\delta\pi(t);t \right],\delta\phi_a(\bx,t) \right] ~,
\quad
\delta\dot\pi_a(\bx,t) = i \left[ \tilde H \left[ \delta\phi(t),\delta\pi(t);t \right],\delta\pi_a(\bx,t) \right] ~.
\label{delta_eom}
\end{align}
However, for our purpose, let us not assume (\ref{comm_relation_0}) for the moment.
Instead we define
\begin{align}
\delta\phi_a &= \sum_{\alpha=1}^N \left[ u_{a}^{(\alpha)} a_{\alpha,\bk} + u_{a}^{(\alpha) *} a_{\alpha,-\bk}^\dagger \right] ~,
\nonumber \\
\delta\pi_a &= \sum_{\alpha=1}^N \left[ w_{a}^{(\alpha)} a_{\alpha,\bk} + w_{a}^{(\alpha) *} a_{\alpha,-\bk}^\dagger \right] ~,
\label{phipi_decomp}
\end{align}
where $a_{\alpha,\bk},a_{\alpha,-\bk}^\dagger$ ($\alpha=1,...,N$) are $N$ sets of the usual creation and annihilation operators, and satisfy the usual commutation relations,
\begin{align}
& [ a_{\alpha,\bp}, a^\dagger_{\beta,-\bq} ] = (2\pi)^3 \delta^3(\bp+\bq) \delta_{\alpha\beta}~,
\nonumber \\
& [ a_{\alpha,\bp}, a_{\beta,-\bq} ] = 0 ~,
\quad
[ a^\dagger_{\alpha,\bp}, a^\dagger_{\beta,-\bq} ] = 0 ~.
\label{commutation_aadagger}
\end{align}
Notice the difference between (\ref{phipi_decomp}) and (\ref{phipi_decomp_usual}) is that in (\ref{phipi_decomp}) each field is decomposed into $N$ sets of creation and annihilation operators.

As the initial conditions, we require that (\ref{phipi_decomp}) satisfy the canonical commutation relation (\ref{comm_relation_0}) at $t_0$, namely
\begin{align}
& \left[ \delta\phi_a(\bx,t_0),\delta\pi_b(\by,t_0)\right] = i\delta_{ab} \delta^3(\bx-\by) ~,
\nonumber \\
& \left[ \delta\phi_a(\bx,t_0),\delta\phi_b(\by,t_0)\right] =
\left[ \delta\pi_a(\bx,t_0),\delta\pi_b(\by,t_0) \right] =0 ~,
\label{comm_relation_t0}
\end{align}
although we still need to show whether (\ref{comm_relation_0}) are satisfied by (\ref{phipi_decomp}) for all $t$. This initial condition translates into a number of conditions for the mode functions $u_a^{(\alpha)}$ and $w_a^{(\alpha)}$.

To give an example without cluttered notation, let us look at the two-field model. The generalization is trivial.
In the two-field model, we need two sets of creation and annihilation operators $a_\bk,a^\dagger_{-\bk}$ and $b_\bk,b^\dagger_{-\bk}$,
\begin{align}
\delta\phi_1 &= u_1 a_\bk + v_1 b_\bk + {\rm c.c.} ~,
\nonumber \\
\delta\phi_2 &= u_2 a_\bk + v_2 b_\bk + {\rm c.c.} ~,
\nonumber \\
\delta\pi_1 &= p_1 a_\bk + q_1 b_\bk + {\rm c.c.} ~,
\nonumber \\
\delta\pi_2 &= p_2 a_\bk + q_2 b_\bk + {\rm c.c.} ~.
\label{phipi_decomp_twofield}
\end{align}
As mentioned, the initial state usually studied are the decoupled initial state, which means that, at the initial time $t_0$, the mode functions in (\ref{phipi_decomp_twofield}) are diagonal,
\bea
v_1=u_2=0 ~, q_1=p_2=0 ~.
\label{Initial_cond_1}
\eea
In such cases, the mode functions $u_1$ and $p_1$, as well as $v_2$ and $q_2$, satisfy the initial commutation condition,
\bea
u_1 p_1^* - u_1^* p_1 =i ~,
\quad
v_2 q_2^* - v_2^* q_2 =i ~.
\label{Initial_cond_2}
\eea

Instead of deriving Eq.~(\ref{delta_eom}) from (\ref{bkgd_eom})-(\ref{comm_relation_0}) as done in the in-in formalism \cite{Weinberg:2005vy,Chen:2010xka,Wang:2013zva}, here we start with this equations but {\em define} the commutators in Eq.~(\ref{delta_eom}) using the definitions in (\ref{commutation_aadagger}). At this moment it is not clear whether the commutation relations (\ref{comm_relation_0}) and (\ref{commutation_aadagger}) are equivalent. Our next step is to construct solutions for $u_{a}^{(\alpha)}$ and $w_{a}^{(\alpha)}$ to achieve this equivalence.

With this definition of commutation relations, we can determine the evolution of $u_{a}^{(\alpha)}$ and $w_{a}^{(\alpha)}$ using (\ref{delta_eom}), which have the following solutions,
\begin{align}
\delta\phi_a(\bx,t) = U^{-1}(t,t_0) \delta\phi_a(\bx,t_0) U(t,t_0) ~,
\quad
\delta\pi_a(\bx,t) = U^{-1}(t,t_0) \delta\pi_a(\bx,t_0) U(t,t_0) ~,
\label{delta_solution}
\end{align}
where $U$ satisfies
\bea
\frac{d}{dt}U(t,t_0) = - i \tilde H \left[ \delta\phi(t_0),\delta\pi(t_0);t \right]  U(t,t_0)
\label{dUdt}
\eea
with the initial condition
\bea
U(t_0,t_0)=1 ~.
\eea
As mentioned,
the initial condition $\delta\phi_a(\bx,t_0)$ and $\delta\pi_a(\bx,t_0)$ are chosen such that (\ref{comm_relation_t0}) is satisfied.
Because of (\ref{delta_solution}), once (\ref{comm_relation_t0}) is satisfied at $t_0$, the commutation relations (\ref{comm_relation_0}) will be satisfied for all $t$.
Therefore, we have ensured the equivalence between the commutation relations (\ref{comm_relation_0}) and (\ref{commutation_aadagger}) for all $t$. The equations (\ref{delta_eom}) defined in two approaches are then equivalent.

To summarize, what we have shown is that the decomposition (\ref{phipi_decomp}) and the corresponding mode functions satisfy the same first-principle quantization conditions and equations of motion as in the in-in formalism, and therefore are equivalent to the in-in formalism in the validity regimes of the two methods.
Now it is easy to see that this decomposition leads to the procedure in Sec.~\ref{Sec:EoM}.
Plugging (\ref{phipi_decomp}) in (\ref{delta_eom}) and matching the coefficients of the $N$ sets of creation and annihilation operators, we can see that we effectively have the same EoM, satisfied by the mode functions $u_{a}^\alpha$, for each set of creation and annihilation operators. The only difference is the initial conditions. For the decoupled initial state, the condition given by (\ref{Initial_cond_1}) and (\ref{Initial_cond_2}) reproduces exactly the condition given in the procedure in Sec.~\ref{Sec:EoM}, (\ref{Initial_W-cond}), (\ref{initial_EoM_1}) and (\ref{initial_EoM_2}), each time we solve the EoM.
Note that to match both sides of (\ref{delta_eom}) using (\ref{phipi_decomp}), the EoM has to be linear. Using (\ref{phipi_decomp}) it is also straightforward to see that the two-point correlation functions are indeed given by (\ref{2pt_EoM}).

The explicit proof also motivates the following considerations.
\begin{itemize}
\item
In the in-in formalism, the perturbative expansion relies on the fact that the bilinear interaction terms in $H_I$ are perturbative with respective to $H_0$. On the other hand, since all we have shown is that the EoM procedure satisfies the same first-principle, this procedure still applies even if $H_I$ is non-perturbative, in which case the linearly coupled EoM can be solved at least numerically.

\item
The most general initial conditions that have to be satisfied are (\ref{comm_relation_t0}). The diagonalized initial commutation relations (\ref{Initial_cond_1}) and (\ref{Initial_cond_2}), which correspond to the case where the fields are initially decoupled, is only a special case. There can be more general cases where the fields are correlated initially. In the EoM approach, this corresponds to non-zero off-diagonal terms in the initial conditions for the mode functions in (\ref{phipi_decomp}), and the rest of the procedure is the same. The in-in formalism can also be modified accordingly.
\end{itemize}
We study these two cases in more details in Sec.~\ref{Sec:EoM_special}.

\section{Examples}
\label{Sec:Examples}
\setcounter{equation}{0}

In this section we explicitly check this equivalence using two non-trivial examples, namely the quasi-single field (QSF) inflation model and the standard clock (SC) model.
In the first example, the potential has an approximate shift symmetry; in the second example, it contains some features.
In each model, we study the cases with different vacuum choices, resulting in either scale-invariant corrections or scale-dependent features in density perturbations.

\subsection{Quasi-single field inflation model}
\label{Sec:QSF}

The QSF inflation model \cite{Chen:2009we,Chen:2009zp,Baumann:2011nk,Sefusatti:2012ye,Norena:2012yi,Chen:2012ge,Noumi:2012vr,Emami:2013lma,Craig:2014rta,Arkani-Hamed:2015bza,Dimastrogiovanni:2015pla} describes a two-field inflation model in which the inflaton is coupled to a massive field with mass of order $H$. The following is the matter sector Lagrangian for such an example \cite{Chen:2009we,Chen:2009zp},
\ba
\CL =
-\frac{1}{2} (R + \sigma)^2 g^{\mu\nu} \partial_\mu \theta \partial_\nu \theta - V_{\rm sr}(\theta)
-\frac{1}{2} g^{\mu\nu} \partial_\mu \sigma \partial_\nu \sigma - V_\sigma(\sigma) ~,
\label{L_m}
\ea
and we assume the following potentials for two fields
\ba
V_{\rm sr}(\theta) = \dfrac12 m^2 R^2 \theta^2 ~,
\\
V_\sigma(\sigma) = \dfrac12 m_0^2  \sigma^2 ~.
\label{V_sigma_QSF}
\ea
See \cite{Chen:2009we,Chen:2009zp,Baumann:2011nk,Sefusatti:2012ye,Norena:2012yi,Chen:2012ge,Noumi:2012vr,Emami:2013lma,Craig:2014rta,Arkani-Hamed:2015bza,Dimastrogiovanni:2015pla} for motivations and generalizations of such models.
The field $\sigma$ has a mass comparable to the Hubble parameter and $\theta$ is the light field driving inflation in a constant turning trajectory. The background evolution in the case of constant turning ($\dot \sigma = 0$) is given by the following equations
\begin{align}
&
3  \Mpl^2 H^2 = \dfrac12 \left(R+\sigma _0\right)^2 \, \dot\theta _0^2+V_{\rm sr}\left(\theta _0\right)+V_{\sigma }\left(\sigma _0\right) ~,
\\
&
\ddot \theta_0 +3 H \dot\theta_0+\dfrac{V_{\rm sr}'\left(\theta _0\right)}{(R+\sigma _0)^2 }=0 ~,
\\
&
V_{\sigma }'\left(\sigma _0\right)=\left(R+\sigma _0\right) \dot\theta _0^2 ~,
\\
&
\epsilon \equiv -\dfrac{\dot H}{H^2} = \dfrac{(R+\sigma _0)^2 \, \dot \theta^2}{2 \Mpl^2 H^2} ~.
\label{QSF_background}
\end{align}
Perturbing the Lagrangian up to second order gives the following Lagrangian for the perturbations
\ba
\CL_2 &\approx&
\frac{a^3}{2} (R + \sigma_0)^2
\left[ \dot{\delta\theta}^2 - \frac{1}{a^2} (\partial_i \delta\theta)^2 \right]
+ \frac{a^3}{2} \dot{\delta\sigma}^2 - \frac{a}{2} (\partial_i \delta\sigma)^2
- \frac{a^3}{2}
V''_\sigma \delta\sigma^2
\nonumber \\
&+& 2a^3 R \dot\theta_0
\delta\sigma \dot{\delta\theta} ~,
\label{L2_bkgd_leading}
\ea
in which we have neglected the mass term for $\delta \theta$ and the second line is the bilinear coupling between two fields. Since $\sigma_0 = {\rm const.}$, we redefined $R$ such that $\sigma_0=0$ in the above equation and hereafter\footnote{In the numerical computations we do not neglect the time dependence of $\sigma_0$ as it may have some evolution at initial times. However, we do neglect the mass term of $\delta \theta$. }. The full equations of motion for the two fields are given by
\begin{align}
&\ddot{\delta\theta}+
 3H
\dot{\delta\theta}
+
 \frac{k^2}{a^2}
\delta\theta
+ 2\left( \dfrac{3H\dot \theta_0+\ddot \theta_0}{R} \right) \delta \sigma + 2 \dfrac{\dot \theta_0}{R} \dot{\delta \sigma}
=0 ~,
\\
&\ddot{\delta \sigma}+3H  \dot{\delta \sigma} +  \left[ V_\sigma'' +\dfrac{k^2}{a^2} \right]\delta \sigma-2 R\dot \theta_0 \dot{\delta \theta}= 0 ~.
\label{QSF_EOM}
\end{align}

As we discussed in the previous sections our proof is not based on the BD initial conditions; so we can assume a more general, but still decoupled, initial condition as follows
\ba
R u_k  = i \dfrac{H\tau}{\sqrt{2k}} \left(C_\theta \, e^{-ik\tau} -D_\theta \, e^{ik\tau} \right) ~,
\quad
v_k=0 ~,
\label{QSF_cond1}
\ea
and
\ba
Ru_k=0 ~,
\quad
v_k  = i \dfrac{H\tau}{\sqrt{2k}} \left(C_\sigma \, e^{-ik\tau} -D_\sigma \, e^{ik\tau} \right) ~,
\label{QSF_cond2}
\ea
where $u_k$ and $v_k$ are the initial mode functions for $\delta \theta$ and $\delta \sigma$ and the above initial conditions have to be set at sufficiently early times, i.e. $k\vert \tau\vert \gg 1$. Initial conditions have to satisfy the commutation relations which are equivalent to requiring
\ba
\vert C_\theta \vert^2 - \vert D_\theta \vert^2 = 1 , \qquad \vert C_\sigma \vert^2 - \vert D_\sigma  \vert^2  = 1.
\ea
Using the above initial conditions one can solve the equations of motion numerically twice following the procedure in Sec.~\ref{Sec:EoM}. The initial condition (\ref{QSF_cond1}) leads to $u_p^{(1)}(t)$, and (\ref{QSF_cond2}) leads to $u_p^{(2)}(t)$.
The two-point correlation function is given by
\begin{align}
\langle \delta\theta_\bp \delta\theta_\bq \rangle =
(2\pi)^3 R^2 \left( |u_p^{(1)}(t)|^2 + |u_p^{(2)}(t)|^2 \right) \delta^3(\bp+\bq) ~.
\end{align}

On the other hand, for the in-in approach, we use the so-called {\it commutator form} because it is more convenient in terms of avoiding some spurious IR divergence in the intermediate steps \cite{Chen:2009we,Chen:2009zp}. For a generic interaction Hamiltonian $H_{I}$ and at the leading order in perturbations we have
\ba
\langle  \delta \theta^2   \rangle \supset
- \int^t_{t_0}  dt_1 \int^{t_1}_{t_0} dt_2  \, \langle [ H_I(t_2) , [H_I(t_1), \delta \theta(t)^2 ]] \rangle ~.
\ea
Assuming the interaction Hamiltonian in the following form
\ba
H_I(t) = C(t) \delta\sigma \dot \delta \theta
\ea
yields
\ba
\langle  \delta \theta^2   \rangle =
-4 \, {\mathrm{Re}} \left[\int^t_{t_0} dt_1 \int^{t_1}_{t_0} dt_2  \, \, C(t_1) C(t_2) \, v(t_2)  \, \dot u(t_2) \, v^*(t_1) \, u^*(t) \left( \dot u(t_1) u^*(t) - u(t) \dot u^*(t_1)  \right)  \right] ~,
\ea
in which $u$ and $v$ are the free field mode functions for $\delta \theta$ and $\delta \sigma$, respectively. For the QSF model we have $H_I = -2a^3 R \dot\theta_0
\delta\sigma \dot{\delta\theta}$.\footnote{Transforming from Lagrangian to Hamiltonian, a small mass term for $\delta \sigma$ will be induced by the bilinear term. We will neglect this extra term.}

\begin{figure}
\includegraphics[scale=.8]{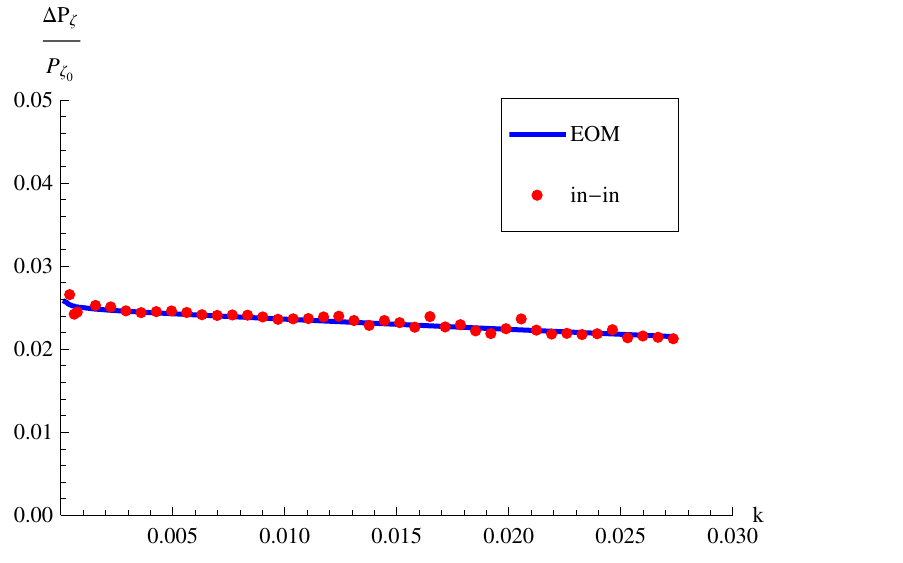}
\includegraphics[scale=.8]{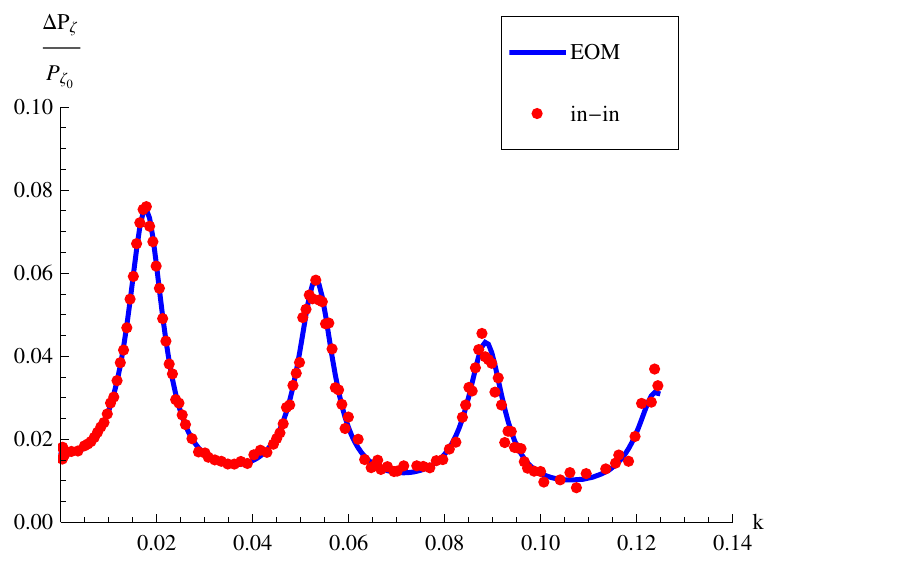}
\caption{\small The correction to the power spectrum for the QSF model in the weak coupling limit. The left (right) panel is the results for the BD (a non-BD) initial conditions. For the non-BD case we have set $D_\theta = D_\sigma = 1/2$ and hence $C_\theta = C_\sigma = \sqrt{5/4}$. For other parameters we have set $R=0.8 \Mpl$, $m_0 \simeq 1.4 H$ and $m \simeq 0.06 H$ which correspond to the coupling $(\dot \theta/H )\simeq 0.004$. }
\label{QSF_weak}
\end{figure}

\begin{figure}
\center
\includegraphics[scale=1]{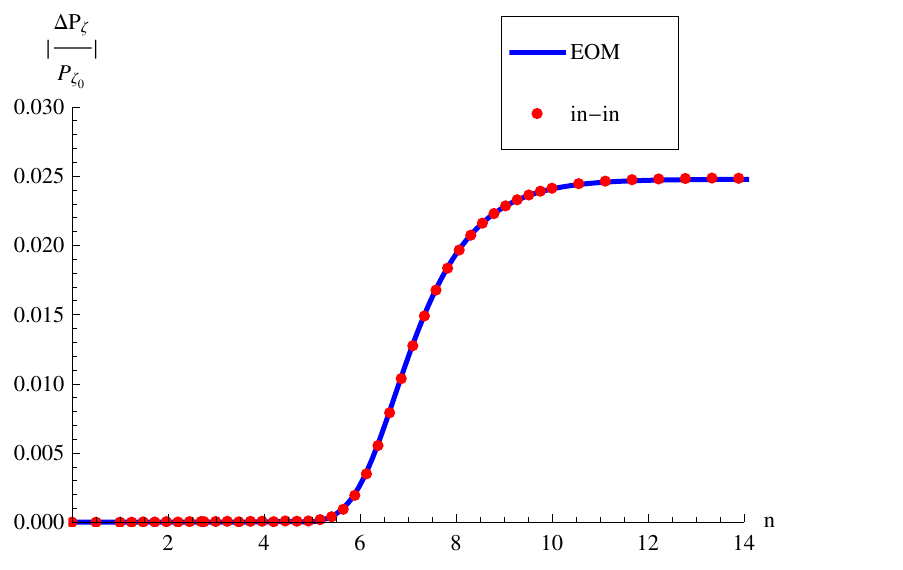}
\caption{\small The time dependence of the correction to power spectrum for the weak coupling case. The horizontal axis is number of efolds counted from the initial time. The parameters are similar to ones in Fig.\ref{QSF_weak}. The mode crosses the horizon around $n \simeq 6$.}
\label{QSF_time}
\end{figure}

In the following figures we plot the fractional correction to the power spectrum of curvature perturbation. The curvature perturbation is related to the light field $\theta$ via
\ba
\zeta = - \dfrac{H}{\dot \theta} \delta \theta.
\ea
In Fig.\ref{QSF_weak} we plot the results for the weak coupling limit in which the two approaches agree pretty well. In Fig.\ref{QSF_time} we show the time dependence of the results from a specific mode in the BD case showing that the equivalence is always valid between two initial and final time slices.

\subsection{Standard clock model}
\label{Sec:SC}

In this subsection, we consider an example in which the potential contain features, namely the standard clock models.
The Lagrangian of the example of the standard clock model in \cite{Chen:2014joa,Chen:2014cwa} is very similar to that of the QSF inflation model. There are two differences in the model setup. First, the mass of the $\sigma$-field is much larger than $H$ in the standard clock model \cite{Chen:2012ge,Pi:2012gf}, so asymptotically the inflaton rolls in the trough of this $\sigma$-potential valley and the effect of the massive field is negligible. Second, an earlier phase that describes how the inflaton settles down to the bottom of this valley is added, during which the massive field plays an important role in density perturbations.
The motivation for this model is given in \cite{Chen:2011zf,Chen:2011tu} and summarized in \cite{Chen:2014cwa}.

So the Lagrangian is still given by (\ref{L_m}) but we modify the potential $V_\sigma$ to
\ba
V_\sigma = V_{\sigma 0} \left[ 1- \exp(-\sigma^2/\sigma_f^2) \right] + \dfrac{1}{2}m_0^2 \sigma^2 ~.
\label{V_sigma}
\ea
The inflaton initially starts somewhere at the plateau of this potential, and then rolls toward the bottom of the valley, oscillates and settles down to an effectively single-field inflation model.
The transition occurs when $\sigma$ field reaches to the critical value $\sigma \simeq \sigma_f$. The background equations can be obtained by varying the Lagrangian \eqref{L_m} which results in
\begin{align}
&
3  \Mpl^2 H^2 = \dfrac12 \left(R+\sigma _0\right)^2 \, \dot\theta _0^2+\dfrac12\dot\sigma _0^2+V_{\rm sr}\left(\theta _0\right)+V_{\sigma }\left(\sigma _0\right) ~,
\\
&
\ddot \theta_0 +3 H \dot\theta_0+\dfrac{2 \dot\theta _0 \dot\sigma _0}{(R+\sigma _0)}+\dfrac{V_{\rm sr}'\left(\theta _0\right)}{(R+\sigma _0)^2 }=0 ~,
\\
&
\ddot \sigma _0+3 H \dot\sigma _0+V_{\sigma }'\left(\sigma _0\right)-\left(R+\sigma _0\right) \dot\theta _0^2=0 ~,
\\
&
\epsilon \equiv -\dfrac{\dot H}{H^2} = \dfrac{(R+\sigma _0)^2 \, \dot \theta^2 + \dot \sigma^2}{2 \Mpl^2 H^2} ~.
\label{epsilon_definition}
\end{align}
For the perturbations we have the free field Lagrangian similar to the one in the QSF model, i.e.~the first line in \eqref{L2_bkgd_leading} (again we neglect the mass term of $\delta \theta$). There are several bilinear terms for this model but for the purpose of this paper we only pick up the following one for simplicity
\ba
\Delta \CL_2 &\approx&
- \frac{a^3}{H} (\sigma_0+R)^2 \dot \sigma_0 \dot\theta_0
\delta\sigma \dot{\delta\theta} ~.
\label{L2_pert_leading}
\ea
The EoM are
\begin{align}
\ddot{\delta\theta} +
\left[ 3H + \frac{2\dot\sigma_0}{R + \sigma_0} \right]
\dot{\delta\theta}
+
 \frac{k^2}{a^2}
\delta\theta \hspace{6.2cm}
\\ \nonumber
 -\left[ (3+\epsilon) \dot \theta_0 \dot \sigma_0+ \ddot{\theta_0}\dot \sigma_0 + \dfrac{2\dot \theta_0 \dot \sigma_0^2}{H\left(R+\sigma _0\right)} +\dfrac{\dot \theta_0 \ddot{\sigma_0}}{H} \right] \delta \sigma
-\dfrac{\dot \theta_0 \dot \sigma_0}{H} \dot{\delta\sigma}
=0 ~,
\\
\ddot{\delta \sigma}+3H  \dot{\delta \sigma} +  \left[ V_\sigma'' +\dfrac{k^2}{a^2} \right]\delta \sigma+  \dfrac{\dot\theta _0   \dot\sigma _0 }{H}\left(R+\sigma _0\right){}^2 \dot{\delta \theta}= 0 ~.
\label{EOM_smallfield}
\end{align}
Finally, to solve the above equations we impose the following generic initial conditions
\ba
\left(R+\sigma_0(\tau)\right) u_k  = i \dfrac{H\tau}{\sqrt{2k}} \left(C_\theta \, e^{-ik\tau} -D_\theta \, e^{ik\tau} \right) ~,
\quad
v_k=0 ~,
\ea
and
\ba
\left(R+\sigma_0(\tau)\right) u_k =0 ~,
\quad
v_k  = i \dfrac{H\tau}{\sqrt{2k}} \left(C_\sigma \, e^{-ik\tau} -D_\sigma \, e^{ik\tau} \right) ~.
\ea
The procedure of solving the EoM, and the formula for the in-in formalism approach, are the same as in Sec.~\ref{Sec:QSF}.

\begin{figure}
\includegraphics[scale=.8]{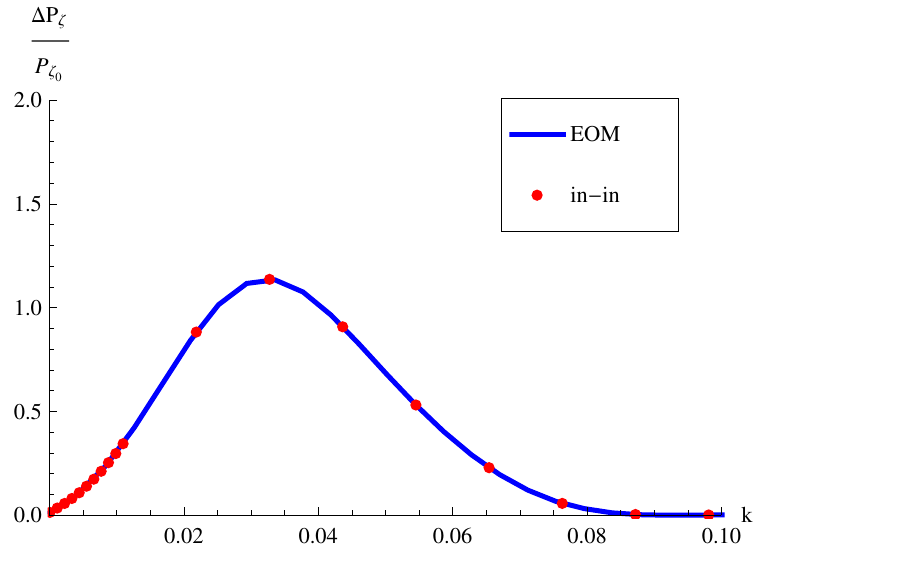}
\includegraphics[scale=.8]{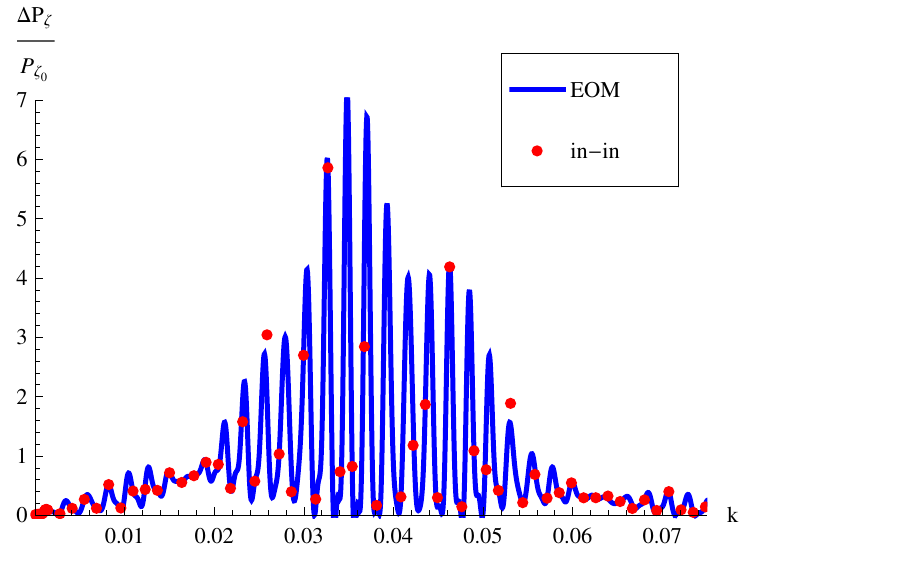}
\caption{\small The correction to the power spectrum for the SC model. The left (right) panel is the results for the BD (a non-BD) initial conditions. For the non-BD case we have set $D_\theta = D_\sigma = 1/2$ and hence $C_\theta = C_\sigma = \sqrt{5/4}$. For other parameters we have set $R=2.7 \Mpl$, $m_0 \simeq 0.7 H$ and $m \simeq 0.1 H$, $\sigma_f \simeq 0.05 \Mpl$, $V_0 \simeq 2.7 \times 10^{-9} \Mpl^4$ and the initial conditions for background fields are tuned such that the transition occurs at $n=6$.}
\label{SC}
\end{figure}

\begin{figure}
\center
\includegraphics[scale=1]{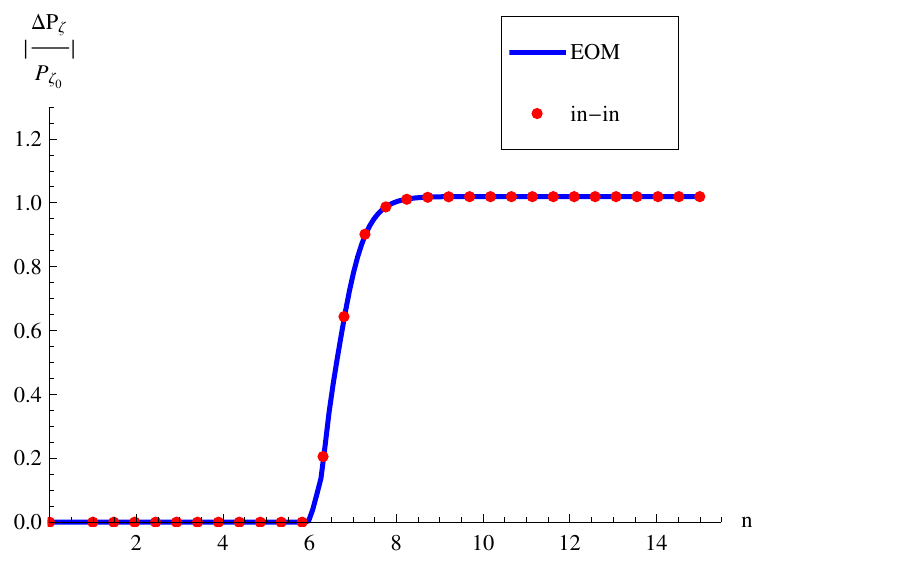}
\caption{\small The time dependence of the correction to power spectrum for the SC model. The horizontal axis is number of efolds counted from the initial time. The parameters are similar to ones in Fig.\ref{SC}. The mode crosses the horizon at $n \simeq 6.4$.}
\label{SC_time}
\end{figure}

In Fig.\ref{SC} we plot the fractional correction to the power spectrum obtained by both approaches for the two cases of BD and non-BD initial conditions.\footnote{The examples used here are different from the best-fit examples in \cite{Chen:2014joa}. Here we used the large field examples where the effect of the 2nd field is more important, although this cases does not fit the data. Our purpose is to show the equivalence between the two approaches.}
In Fig.\ref{SC_time}, we show the time-dependence of one of the modes in the BD case. We see that two approaches match very well.\footnote{We observe that even for a large correction the two approaches match pretty well. It seems even for this seemingly non-perturbative case the higher order corrections in the in-in formalism are still negligible. 
We currently do not understand why this is the case, but we expect generally this would not be true
 (see Sec.~\ref{Sec:Strong_coupling}).}

\section{Several extensions}
\label{Sec:EoM_special}
\setcounter{equation}{0}
The procedures in the two approaches reviewed in Sec.~\ref{Sec:EoM} and Sec.~\ref{Sec:inin} are commonly used in the literature.
In this section, motivated by the explicit proof of the equivalence between the two approaches presented in Sec.~\ref{Sec:Proof}, we discuss several extensions of these procedures that may be applied to new categories of models.

\subsection{Density perturbations with strong coupling}
\label{Sec:Strong_coupling}

As we mentioned at the end of Sec.~\ref{Sec:Proof}, the validity of the EoM approach does not rely on the perturbative condition. Therefore, solving the linearly coupled differential equations in the EoM approach produces the full tree-level non-perturbative two-point correlation function. Formally in terms of the in-in formalism, this would correspond to a non-perturbative re-summation of all the tree-level diagrams for the two-point correlation function.

We emphasize that this is of course not the full non-perturbative result because the loop diagrams are not included; nonetheless it is an interesting subset. Another limitation is that this procedure only applies to the tree-level two-point correlation functions due to the linearity condition in the EoM approach assumed in the proof in Sec.~\ref{Sec:Proof}. Possible generalization would be very interesting.

\begin{figure}
\includegraphics[scale=.8]{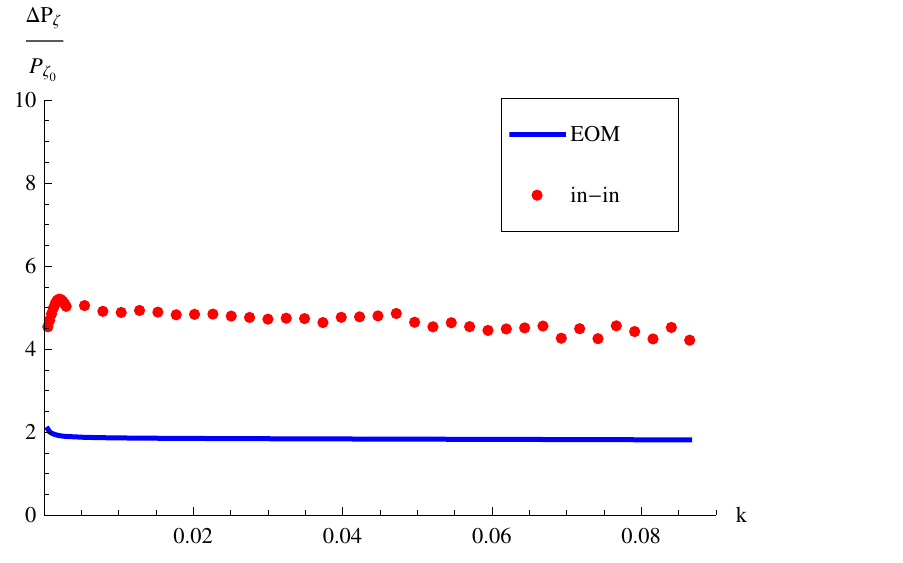}
\includegraphics[scale=.8]{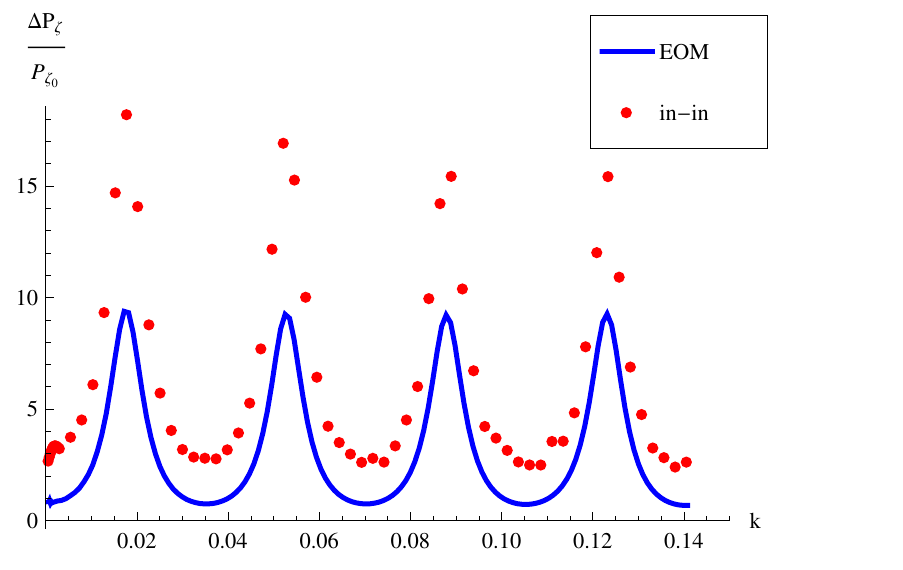}
\caption{\small The correction to the power spectrum for the QSF model in the strong coupling limit. The left (right) panel is the results for the BD (non-BD) initial conditions. For the non-BD case we have set $D_\theta = D_\sigma = 1/2$ and hence $C_\theta = C_\sigma = \sqrt{5/4}$. Other parameters are tuned as follows: $R=0.05 \Mpl$, $m_0 \simeq 1.4 H$ and $m \simeq 0.18 H$ which correspond to the coupling $(\dot \theta/H) \simeq 8.64$. In both figures, the blue lines are the EoM results, and the red dots are the first order result from the in-in formalism.}
\label{QSF_strong}
\end{figure}

To study this in an example, we note that, in the QSF inflation model, the coupling between the inflaton and the massive field $\sigma$ is of order $\dot\theta/H$. In the previous examples, this coupling is taken to be small so the correction to the leading power spectrum is small. For large $\dot\theta/H$, the perturbative expansion in the in-in formalism breaks down, while the numerical computation in the EoM approach is essentially unchanged. In Fig.\ref{QSF_strong} we plot the results for such a strong coupling case. We see large mismatch between the result from the EoM approach and that from the first order term in the in-in approach. This shows that the higher order corrections in the in-in formalism are non-negligible.

It would be interesting to see if the EoM may be solved analytically, or if all the tree-level diagrams in the in-in formalism may be re-summed. We leave these for future investigation.

\subsection{Correlated initial states}
\label{Sec:correlated_initial}

In this subsection we consider a more general initial condition in which the two fields are correlated. As we already discussed, we only need to satisfy initial commutation relations and the proof works even if the fields are correlated initially. We show here the consistency of the two approaches in this general case in an explicit example.

There may be variety of reasons that the fields are initially correlated, for example, due to some interactions in a pre-existing phase. See \cite{Albrecht:2014aga} for an example. Our approach here is phenomenological. We parameterize as many such possibilities as we can and do not attempt to construct concrete models. We vary these parameters and study their effects in the density perturbations. We also use both the EoM and in-in approaches to compute the same problems and demonstrate their equivalence.

We start from the following two correlated fields and their associate conjugate momenta
\ba
\delta \phi = u_1 a + v_1 b +c.c.
\\
\pi_\phi = p_1 a + q_1 b + c.c.
\\
\delta \sigma = u_2 a + v_2 b+c.c.
\\
\pi_\sigma = p_2 a + q_2 b + c.c.
\ea
These are $8$ complex parameters or equivalently $16$ real parameters.  We need to satisfy the following commutation relations
\bea
[\delta \phi , \pi_\phi ]=  \{u_1,p_1 \} + \{ v_1, q_1 \} = i
\label{th_pth}
\\
\left[\delta \sigma , \pi_\sigma \right]=\{u_2,p_2 \} + \{ v_2, q_2 \} = i
\label{s_ps}
\\
\left[\delta \phi , \delta \sigma \right]=  \{u_1,u_2 \} + \{ v_1, v_2 \} = 0
\label{th_s}
\\
\left[\delta \phi , \pi_\sigma \right]=  \{u_1,p_2 \} + \{ v_1, q_2 \} = 0
\label{th_ps}
\\
\left[\delta \sigma , \pi_\phi \right]=  \{u_2,p_1 \} + \{ v_2, q_1 \} = 0
\label{s_pth}
\\
\left[\pi_\phi , \pi_\sigma \right]=  \{p_1,p_2 \} + \{ q_1, q_2 \} = 0
\label{pth_ps}
\eea
in which we defined
\ba
\{p,q \} \equiv p q^* - p^* q.
\ea
Note that the above six equations are constraints for the imaginary part only, as the real part of the combination in the l.h.s vanishes automatically. Hence, in the most general case, we have $6$ constraint for $16$ parameters and as a result $16-6=10$ real independent parameters.

To go further and have more control on the parameters we can restrict ourselves to the case in which the following parameters satisfy the usual commutation relations, same as those in the initially uncorrelated case,
\ba
\{u_1,p_1 \} = \{v_2, q_2 \}  = i .
\label{uncorrelated}
\ea
Plugging these into \eqref{th_pth} and \eqref{s_ps} yields
\ba
\{v_1 , q_1 \} = \{u_2 , p_2 \} = 0.
\label{v1q1}
\ea
This splitting of equations gives two more constraints so it appears that we have $8$ free real parameters left. However, after this splitting, one can check that one of the equations would be redundant and can be reproduced from the others. So we actually have $9$ independent parameters. We can further parameterize by noting that \eqref{v1q1} implies that the combinations $v_1 q_1^*$ as well as $u_2 p_2^*$ are both real. We thus can define
% \ba
% v_1 q_1^* = \tilde{R_1}
% \\
% u_2 p_2^* = \tilde{R_2}
% \ea
% or equivalently
\ba
q_1= R_1 v_1
\\
p_2 = R_2 u_2
\ea
in which $R_i$
% and $\tilde{R_i}$
are real parameters. Note that this last step is just a new way of parameterization and does not restrict the independency of parameters. Similarly we can rewrite \eqref{th_s} and \eqref{th_ps} by
\ba
u_1 u_2^* + v_1 v_2^* &=& A \left( \vert u_1 \vert^2 + \vert v_2 \vert^2 \right)
% \\
% u_1 p_2^*+v_1 q_2^* &=& R_3
\ea
in which $A$
% and $R_3$ are
is real and the factor in the bracket in the r.h.s makes $A$ dimensionless. We will see soon that the strength of the initial correlation is controlled by $A$. After some algebra we can show that $v_1,q_1,u_2$ and $p_2$ can be expressed as secondary parameters in terms of $u_1,p_1,v_2,q_2,R_1,R_2$ and $A$,
\begin{align}
  v_1 & = \frac
        {A(q_2 - R_2 v_2)(|u_1|^2+|v_2|^2)}
        {p_1 u_1^* + q_2^* v_2 - R_1 |u_1|^2 - R_2 |v_2|^2}~,
        \label{v1}
  \\
  u_2 & = \frac
        {A(p_1 - R_1 u_1)(|u_1|^2+|v_2|^2)}
        {p_1 u_1^* + q_2^* v_2 - R_1 |u_1|^2 - R_2 |v_2|^2}~.
        \label{u2}
\end{align}
% \ba
% v_1 &=& \dfrac{R_3 -R_2 A \left( \vert u_1 \vert^2 + \vert v_2 \vert^2 \right)}{q_2^* -v_2^* R_2 }
% \\
% u_2 &=& \dfrac{A q_2 \left( \vert u_1 \vert^2 + \vert v_2 \vert^2 \right)  - R_3 v_2}{u_1^*(q_2 -v_2 R_2 )}
% \\
% R_3 &=&  A \left( \vert u_1 \vert^2 + \vert v_2 \vert^2 \right) \dfrac{\vert q_2 \vert^2 -R_1 R_2 \vert u_1 \vert^2 + R_2 (u_1^* p_1 - v_2^* q_2)}{v_2 q_2^* + u_1^* p_1 - R_2 \vert v_2 \vert^2 - R_1 \vert u_1 \vert^2 }
% \ea
% As a consistency check one can prove that $R_3$ is real.

As for the parameters in Eq.~\eqref{uncorrelated} we can simply write them as a combination of positive and negative frequency of vacuum in Minkowski space to make sure that they satisfy their own commutation relation as well. That is
\ba
u_1  \sim i \dfrac{H\tau}{\sqrt{2k}} \left(C_\phi \, e^{-ik\tau} -D_\phi \, e^{ik\tau} \right)
\\
v_2  \sim i \dfrac{H\tau}{\sqrt{2k}} \left(C_\sigma \, e^{-ik\tau} -D_\sigma \, e^{ik\tau} \right).
\ea
and their corresponding conjugate momenta. To satisfy commutation relations we require
\ba
\vert C_\phi \vert^2 - \vert D_\phi \vert^2 = 1 , \qquad \vert C_\sigma \vert^2 - \vert D_\sigma  \vert^2  = 1.
\label{norm}
\ea
To summarize we have nine real free quantities parameterized as follows
\ba
C_\phi , \, D_\phi , \, C_\sigma , \, D_\sigma , \, R_1, \, R_2, \, A.
\ea
The first four parameters are complex numbers for the negative and positive frequency vacuum which have to satisfy \eqref{norm}. The remaining three parameters are real and are responsible for correlation. Turning off these parameters turns off the correlation as well. Note that $A$ is multiplied to all correlation terms. That is, setting $A =0$ turns off all the correlation terms.

So far we have parameterized a subclass of models with correlated initial states, in the following we check the consistency of two approaches using the above parameterization.
%\\\\
%%Here is a side comment regarding the overall phase difference between initial conditions: It turns out that the overall phase difference between initial condition of $\delta \sigma$ and $\delta \theta$ is not an observable, at least for our specific way of parametrization. To see this let us look at the in-in integral at leading order in the initial correlated case:
%\ba
%\langle \delta \theta^2 \rangle \supset
%-4 \,{\mathrm{Im}}
%\left[ \int dt_1 C(t_1)
%\biggl( u_2(t_1) u_1^*(t) + v_2(t_1) v_1^*(t) \biggl)
%\times
%\biggl( \dot u_1(t_1) u_1^*(t) +  \dot v_1(t_1) v_1^*(t) \biggl)
% \right]
%\ea
%

There is an interesting difference between the correlated and un-correlated initial conditions in the in-in expansion. In the case of correlated initial condition we need only one interaction Hamiltonian to obtain non-zero two-point function while in the case of un-correlated we need two. This is a direct consequence of the fact that in the former case two fields are correlated (even in the absence of interaction) and hence their contraction is non-vanishing. As a result, if the strength of the  interaction Hamiltonian is typically $\alpha$, the leading order correction to the zeroth order power spectrum is at the order of $\alpha$ in the correlated case and at the order of $\alpha^2$ in the un-correlated case.

If the fields are initially correlated, the leading order correction to the power spectrum is given by the following expression
\bea
\langle   \delta \theta^2 \rangle
&\supset & i \langle \Omega \vert  \int dt_1 [H(t_1),\delta \theta(t)^2] \vert \Omega \rangle
\\  \nonumber
&=&
-4 i \, {\mathrm{Im}} \biggl\{ \int dt_1 C(t_1)  \left[u_2(t_1) u_1^*(t) + v_2(t_1) v_1^*(t) \right]
\left[ \dot u_1(t_1) u_1^*(t) + \dot v_1(t_1) v_1^*(t)  \right]
 \biggl\}
\label{2pt_correlated_initial}
\eea
in which we assumed $H_I(t)=C(t) \delta \sigma \dot{\delta \theta }$.

The common phase of the two fields, $\delta\phi$ and $\delta\sigma$, is not observable. At least for the case considered here, the relative phase between the two fields is not allowed. This is implied by the relations (\ref{v1}-\ref{u2}) where we have to rescale $\delta\phi$ and $\delta\sigma$ by the same phase.
As a result of this observation we can omit two independent but irrelevant parameters as overall phases for both fields. Hence we effectively have seven real independent parameters.

\begin{figure}
\center
\includegraphics[scale=.95]{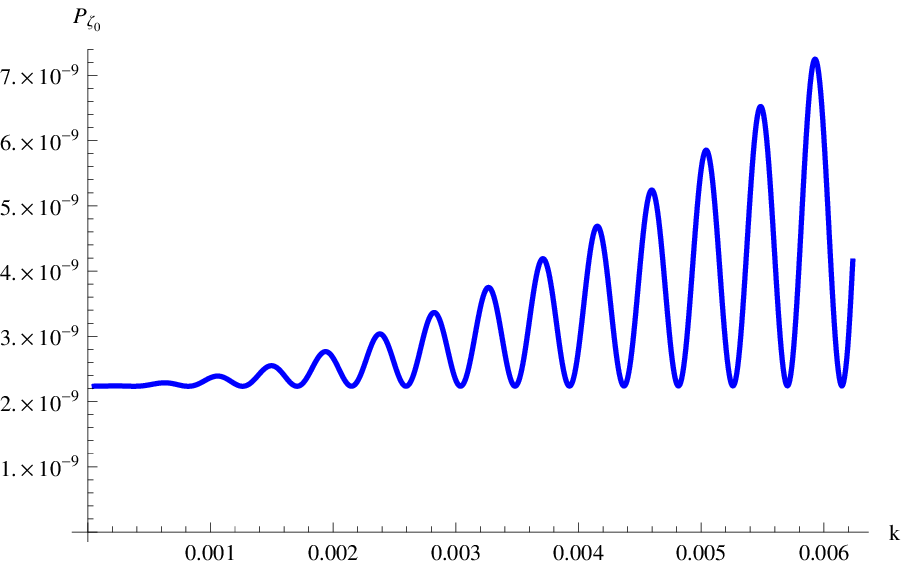}
\includegraphics[scale=1]{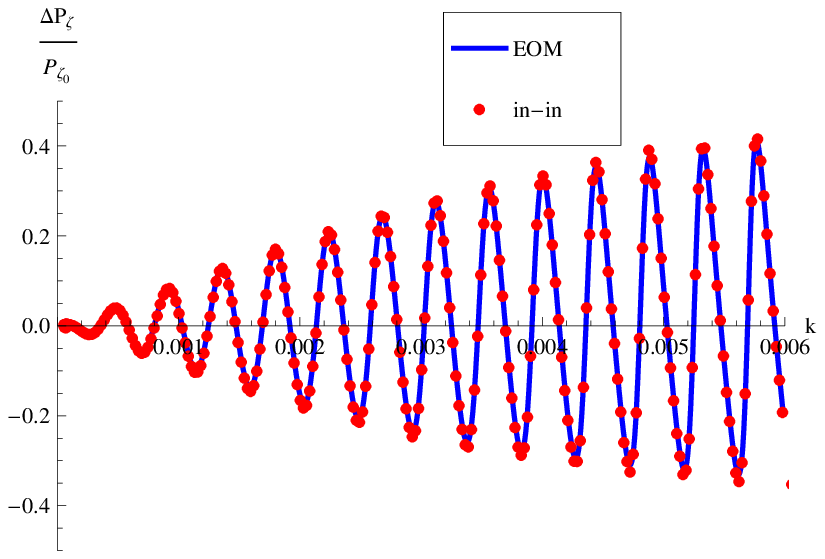}
\caption{\small The zeroth order (top) as well as the correction to the power spectrum due to the coupling (bottom) in QSF model in the case of initially correlated case. The parameters are the following: $R=0.5 \Mpl$, $m_0\simeq 1.4 H$, $m \simeq 0.016 H$, $D_\phi =D_\sigma =0$, $R_1=0$, $R_2=0$ and $A=10^{-3}$. }
\label{ent31}
\end{figure}

\begin{figure}
\center
\includegraphics[scale=.9]{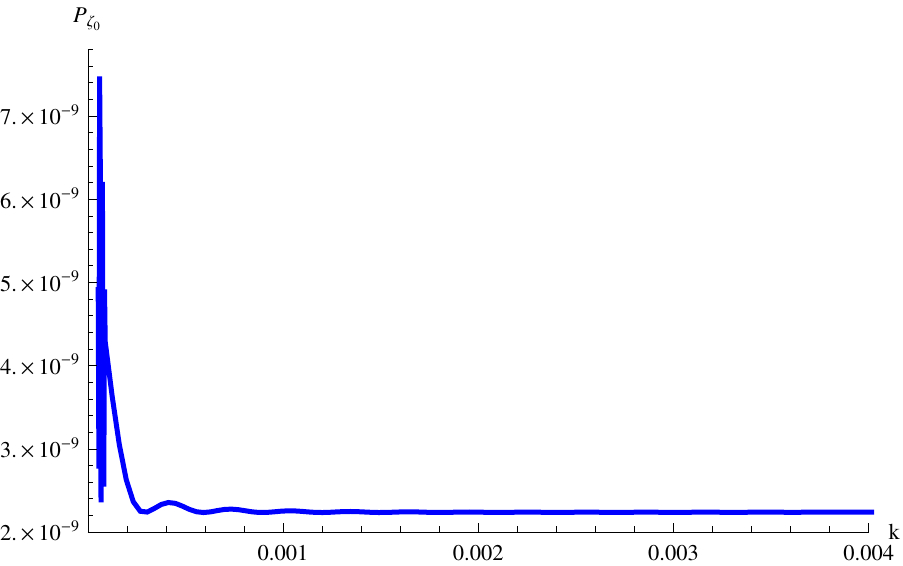}
\includegraphics[scale=1]{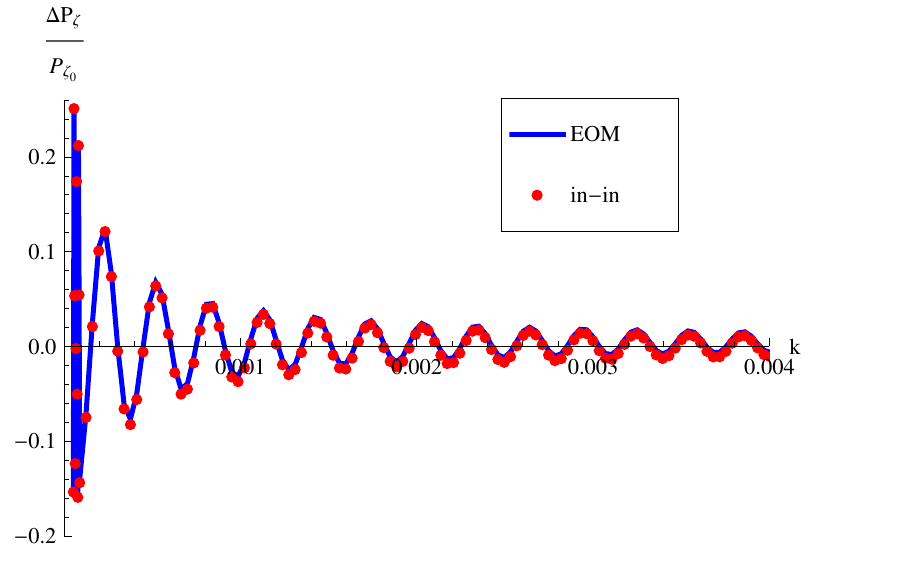}
\caption{\small The same plots as in Fig.\ref{ent31} but with  $R_1=1$, $R_2=1$. Other parameters are as follows: $R=0.5 \Mpl$, $m_0\simeq 1.4 H$, $m \simeq 0.016 H$, $D_\phi =D_\sigma =0$, and $A=10^{-4}$. }
\label{ent32}
\end{figure}

\begin{figure}
\center
\includegraphics[scale=.84]{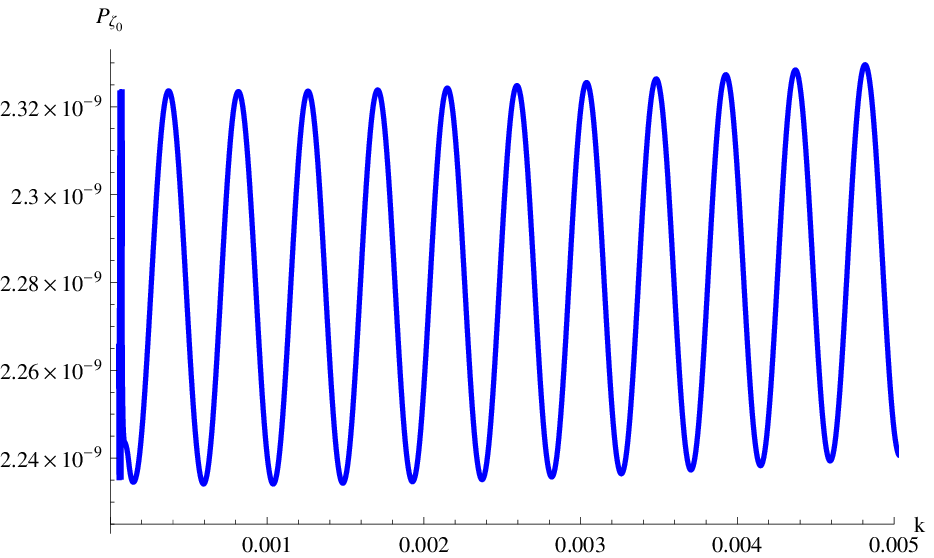}
\includegraphics[scale=1]{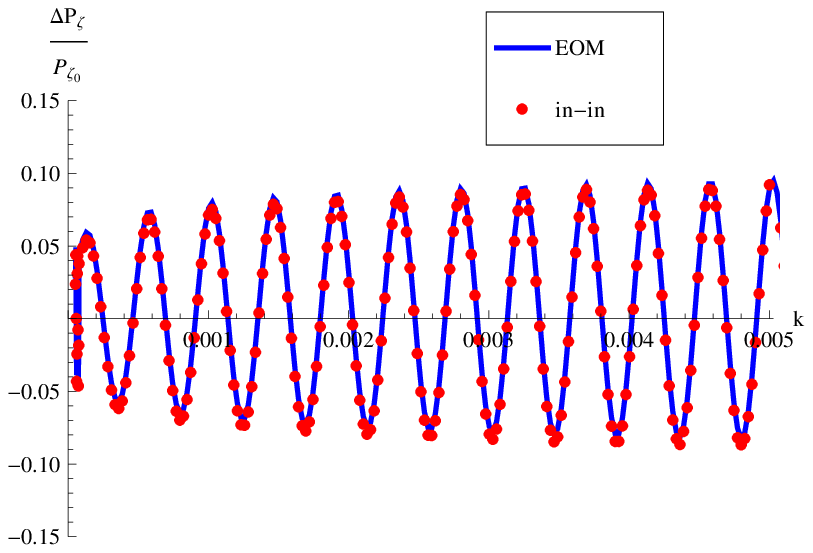}
\caption{\small The same plots as in Figs.\ref{ent31} and \ref{ent32} but with $R_1=1$, $R_2=0$. Other parameters are fixed by the following values: $R=0.5 \Mpl$, $m_0\simeq 1.4 H$, $m \simeq 0.016 H$, $D_\phi =D_\sigma =0$ and $A=0.1$. }
\label{ent}
\end{figure}

Another important consequence of initial correlation is that the free field results, i.e. when the field coupling is set to zero, is also modified as we need to solve the decoupled equations twice with different initial conditions and the final result is the combination of both solutions. Hence in the following figures we also present the free field results besides the correction to the free field due to the field coupling. Figs.~\ref{ent31}, \ref{ent32} and \ref{ent} show the results for the QSF model in three different set of correlation parameters.

Although the strength of the initial correlation is controlled by the parameters $A$, the overall behavior of the final results is also quite sensitive to $R_1$ and $R_2$ as well. For example,
if $\vert R_1 \vert  \ll 1 $ and $ \vert R_2 \vert \ll 1$, the power spectrum grows rapidly as a function of scale (Fig.\ref{ent31}).
If $\vert R_1 \vert  \gtrsim 1$ and $\vert R_2 \vert  \gtrsim 1$, there are some large spikes at largest scales in the power spectrum (Fig.\ref{ent32}). Finally,
if either  ($\vert R_1 \vert  \gtrsim 1$ and $\vert R_2 \vert  \ll 1$)  or  ($\vert R_1 \vert  \ll 1$ and $\vert R_2 \vert  \gtrsim 1$)  then the power spectrum is oscillatory around an averaged value. In this case the amplitude and frequency of oscillations seem to be unchanged throughout scales (Fig.\ref{ent}).
The above observations show that the initial correlation is a rich phenomenon and deserves further investigation which is beyond the scope of this paper. We also note that the parameterization we have chosen is not the most general one and we have one real parameter less than the most general case.

Higher order correlation functions with such initial conditions can also be computed in the in-in formalism by extending the procedure reviewed in Sec.~\ref{Sec:inin}. As we have done here, we replace the decomposition (\ref{phipi_decomp_usual}) with (\ref{phipi_decomp}). The contractions between different fields are then modified accordingly with the definition of (\ref{phipi_decomp}). With these modifications, Eq.~\eqref{in-in_formalism} can be straightforwardly applied to perturbatively compute not only the power spectrum as we did here, but also any higher order correlation functions.

\section{Conclusion and discussions}
\label{Sec:conclusion}

To conclude, in this work we have shown the equivalence between the EoM and in-in approach for cosmological linear perturbations. A few non-trivial examples are provided.

Inspired by the proof, we extended the formalisms of both approaches beyond the types of models that are usually considered in the literature. This deserves further study.

\begin{enumerate}
  \item Non-perturbative linear perturbation. We used the transfer vertex in QSF inflation as an example. In the parameter space that we have studied, the correction from the transfer vertex in the EoM approach is smaller than that of the first order result in the in-in approach. It is interesting to carry out a more complete survey of parameter space. More importantly, since the more characteristic property of QSF inflation model, i.e.~the intermediate shape non-Gaussianities, shows up in the level of non-linear perturbation, it is interesting to seek a prescription that treats the in-in and EoM approaches complementarily: Use the EoM approach to determine the transfer vertex, and then combine with in-in to calculate the three-point interaction. We hope to study it in a future work.
  \item Multi-field inflation with correlated initial states. We have parameterized the initial state and studied simple examples among all possibilities. It is interesting to further explore the effects in different parameter space, and to relate the current study to inflationary model buildings and figure out scenarios in which the initial correlation show up.
\end{enumerate}

It is also interesting to seek for a general prescription of EoM approach at the nonlinear level for multi-field inflation. See e.g. \cite{Dias:2015rca} for linear order. 

\section*{Acknowledgments}

We thank A.~A.~Abolhasani, H.~Firouzjahi, X.~Gao, D.~Langlois, S.~Mizuno, M.~Noorbala and R.~Saito for helpful discussions.
XC and MHN are supported in part by the NSF grant PHY-1417421. YW is supported by the CRF Grants of the Government of the Hong Kong SAR under HUKST4/CRF/13G.

\end{spacing}

\end{document}